\documentclass[english,twocolumn,prx,amssymb,amsmath,superscriptaddress]{revtex4}
\usepackage[latin9]{inputenc}
\usepackage{graphicx,color}
\usepackage{color}
\usepackage[bookmarks]{hyperref}
\usepackage{babel}
\usepackage{braket}

\newcommand{\be}{\begin{equation}}
\newcommand{\ee}{\end{equation}}
\newcommand{\bea}{\begin{eqnarray}}
\newcommand{\eea}{\end{eqnarray}}

\definecolor{mygreen}{rgb}{0,0.5,0}
\definecolor{myblue}{rgb}{0,0,0.75}
\definecolor{mymagenta}{cmyk}{0,1,0,0.12}

\usepackage{umoline}

\begin{document}

\title{Magnetic-field induced spiral order in the electric polarization}

\author{Pei Wang}
\affiliation{Department of Physics, Zhejiang Normal University, Jinhua 321004, China}
\email{wangpei@zjnu.cn}

\author{You-Quan Li}

\affiliation{School of Physics, Zhejiang University, Yuhangtang Road 866, Hangzhou 310058, China}
\affiliation{Chern Institute of Mathematics, Nankai University, Weijin Road 94,
Tianjin 300071, People's Republic of China}
\email{yqli@zju.edu.cn}

\date{\today}

\begin{abstract}
We present a phenomenological model for magnetoelectricity in multiferroic materials.
The distinctive feature of the model is a two-component complex order parameter
that encodes the electric polarization, along with a direct coupling
between the polarization and magnetic field.
Our model effectively elucidates that a sufficiently strong magnetic field can destroy electric polarization.
Furthermore, the transition field strength diminishes with rising temperature, following a power-law relation with the exponent being precisely worked out.
At lower field strength, the electric polarization takes a spiral order in the magnetic field, with the spiral wavelength inversely proportional to the magnetic field strength.
We anticipate these predictions can be experimentally tested in future studies on multiferroic materials.
\end{abstract}

\maketitle

%\date{\today}

\section{Introduction}

Multiferroic materials exhibiting a magnetoelectric effect continuously capture theoretical and practical interests~\cite{Hur04,Dong15,Spaldin19}
because they offer novel avenues for achieving electric-field control
of magnetism or magnetic-field control of polarization~\cite{Fiebig05,Fiebig16}.
The microscopic mechanisms underlying magnetoelectric
coupling remain under debate, with the Dzyaloshinksii-Moriya
interaction~\cite{Sergienko06} or spin-lattice
coupling~\cite{Sergienko06b} generally considered pivotal. At the same time,
phenomenological descriptions of magnetoelectricity are popular
in practice, because they need less computational effort and have general
applicability even in the presence of inhomogeneous external fields.
Some experimental observations, such as those in $\text{GdFeO}_3$~\cite{Tokunaga09},
reveal that a strong magnetic field can destroy electric polarization
in materials exhibiting spontaneous polarization below
the antiferromagnetic ordering temperature.
Recent studies also suggest that electric polarization in heterostructures
and superlattices may exhibit inhomogeneity, potentially
featuring domain walls~\cite{Hong21} or topological structures like meron~\cite{Wang20},
vortex-antivortex arrays~\cite{Yadav16,Abid21} or
polar-skyrmion bubbles~\cite{Das19}.

Phenomenological theories usually take electric polarization
$\textbf{P}$ as an order parameter, and include
higher-order couplings~\cite{Zhang08,Winkler20,Ying22} between $\textbf{P}$ and
magnetic moment $\textbf{M}$. For instance, models featuring cubic coupling have been
proposed~\cite{Mostovoy05}. Because $\textbf{P}$
and $\textbf{M}$ further interact with electric and magnetic fields, respectively,
via terms like $\textbf{P}\cdot \textbf{E} $ and
$\textbf{M}\cdot \textbf{B}$, this explains the magnetoelectric
effect - indirect influence of $\textbf{B}$ on $\textbf{P}$.
However, choosing $\textbf{P}$ as an order parameter
neglects the quantum coherent origin of electric polarization.
Therefore, different phenomenological descriptions are worth exploring.
Recently, we proposed a nonabelian Ginzburg-Landau
theory~\cite{YQLi} as a phenomenological model for magnetoelectricity.
The model's distinctive feature is a two-component complex
order parameter for electric polarization, reflecting the
coherent nature of charges. The magnetic field is directly
coupled to this complex order parameter, explaining its
impact on polarization.

In our initial proposal, parameters were temperature-independent.
In this paper, we develop the model by introducing temperature-dependent parameters.
Consequently, the free energy expression aligns with the
orthodox Landau-Devonshire energy~\cite{Devonshire} in the homogeneous
limit.
To demonstrate the model's predictive power, we
explore the influence of an external magnetic field on
electric polarization.
Our model successfully reproduces the destruction of electric polarization by magnetic field,
and also predicts a topological structure in the polarization.
We anticipate our predictions can be experimentally tested in the future.

The paper is organized as follows:
In Sec.\ref{sec:model},
we introduce a phenomenological expression of free energy.
In Sec.\ref{sec:int},
we minimize the free energy to obtain the
polarization in the spatially uniform approximation.
Furthermore, as rigorous formulation of our theory,
the consideration of spatial variance in the polarization is presented in Sec.\ref{sec:spiral}.
Finally, we summary our findings in Sec.\ref{sec:summ}.

\section{Model and method}
\label{sec:model}

We consider a multiferroic material
which is placed in a homogeneous external magnetic field ($\textbf{B}\neq 0$)
without external electric field ($\textbf{E}=0$).
The phenomenological free energy density is given by
\begin{equation}
\begin{split}\label{eq:freeF}
\mathcal{F} = & \frac{1}{2}  \left|
\nabla_i \Psi + \kappa \mathcal{A}_i \Psi\right|^2
+ \left(T-T_0\right) \frac{\alpha}{2} \left| \Psi\right|^4 + \frac{\beta}{4}
\left| \Psi\right|^8 ,
\end{split}
\end{equation}
where $\Psi = \left(\psi_1, \psi_2\right)^T$ is a two-component
complex order parameter that represents the electric polarization,
and $\mathcal{A}_i$ with $i=x,y,z$ are three $2$-by-$2$ matrices
that form
the SU(2) expression of the electromagnetic field.
$T$ is the temperature of the system
and $T_0$ denotes the ferroelectric transition temperature at $\textbf{B}=0$.
Note that the $T_0$ is a defining property of a multiferroic material.
Here $\alpha$, $\beta$ and $\kappa$ are three phenomenological
parameters.
Especially, $\kappa$ denotes the strength of magnetoelectric coupling.

The expression of equation~\eqref{eq:freeF} is a SU(2) generalization of the well known
Ginzburg-Landau theory~\cite{Ginzburg50}
that was proposed for describing superconductors in the presence of electromagnetic fields.
In Eq.~\eqref{eq:freeF}, we express the three components of the polarization vector $\textbf{P}$
in real space in terms of a two-component $\Psi$ in complex space.
Because there is a homomorphism between SU(2) and SO(3),
the connection between them are defined by
$P_i = \Psi^\dag {\sigma_i} \Psi$, where ${ \sigma_i}$
with $i= x, y, z$ are the Pauli matrices, and $P_i$ are the
components of $\textbf{P}$.
Note that a spin SU(2) together with charge U(1) was considered~\cite{Frohlich93}
in an understanding of the fractional quantum Hall effects,
and later considered~\cite{Zaanen08} in exploring spin superfluidity.
It is easy to see $\left| \Psi\right|^2 = P$,
where $P \equiv \left| \textbf{P}\right|$ is the module of the polarization vector.
With this relation, the potential term in Eq.~\eqref{eq:freeF} can be reexpressed as
\begin{equation}
\left(T-T_0\right) \frac{\alpha}{2}
\left| \Psi\right|^4 + \frac{\beta}{4} \left| \Psi\right|^8 = \left(T-T_0\right) \frac{\alpha}{2}P^2
+ \frac{\beta}{4} P^4.
\end{equation}
The potential term here
is equivalent to that in the traditional Devonshire theory~\cite{Devonshire},
which differs from that in our previous paper~\cite{YQLi}.
The current potential term is more realistic when considering temperature dependence of a genuine system.

In Eq.~\eqref{eq:freeF}, $\mathcal{A}_i$ are called the SU(2) gauge
fields.
The idea of representing electromagnetic field in terms of $\mathcal{A}_i$
is inspired by the following fact.
The force that a moving electric dipole undergoes in an electromagnetic field
can be compactly expressed as the product between the dipole four-current
and the SU(2) field strength~\cite{YQLi}.
Here we follow Ref.~[\onlinecite{YQLi}]
and define $\mathcal{A}_i$ as
$\mathcal{A}_i = \frac{i}{\sqrt{2}} \sum_{j,k} \epsilon_{ijk} B_j \sigma_k$,
where $\epsilon_{ijk}$ is the anti-symmetric Levi-Civita tensor.
Consequently, these $\mathcal{A}_i$ are given by
\begin{equation}
\begin{split}
\mathcal{A}_x = &\frac{1}{\sqrt{2}}
  \left(\begin{array}{cc} iB_y & -B_z  \\ B_z & -iB_y
   \end{array}\right)
    \\[2mm]
\mathcal{A}_y = & \frac{1}{\sqrt{2}}
  \left(\begin{array}{cc} -iB_x & iB_z \\ iB_z & iB_x \end{array}\right)
    \\[2mm]
\mathcal{A}_z = & \frac{1}{\sqrt{2}}
  \left(\begin{array}{cc} 0 & B_x -iB_y \\ -B_x -iB_y & 0 \end{array}\right).
\end{split}
\end{equation}
As considering a uniform magnetic field,
we can choose the direction of $\textbf{B}$ as the $y$-axis
without loss of generality,
which means $B_y = B\neq 0$ but $B_x=B_z=0$.

Equation~\eqref{eq:freeF} is a Ginzburg-Landau energy density.
Strictly speaking,
one needs carry out a path integral of $\mathrm{e}^{-\int \mathrm{d}^3\textbf{r} \mathcal{F}}$
over $\Psi(\textbf{r})$ for obtaining its average.
But in this paper, as a first step, we minimize the free energy $-\int \mathrm{d}^3\textbf{r} \mathcal{F}$
to obtain $\Psi(\textbf{r})$.
This approach can be regarded  as an approximation by dropping the fluctuation completely.
Such a simplified treatment has been frequently employed
in the study of Landau-type models,
before the more complicated field-theoretical method is involved.

\section{Homogeneous approximation}
\label{sec:int}

\begin{figure}[tbp]
\includegraphics[width=1.0\linewidth]{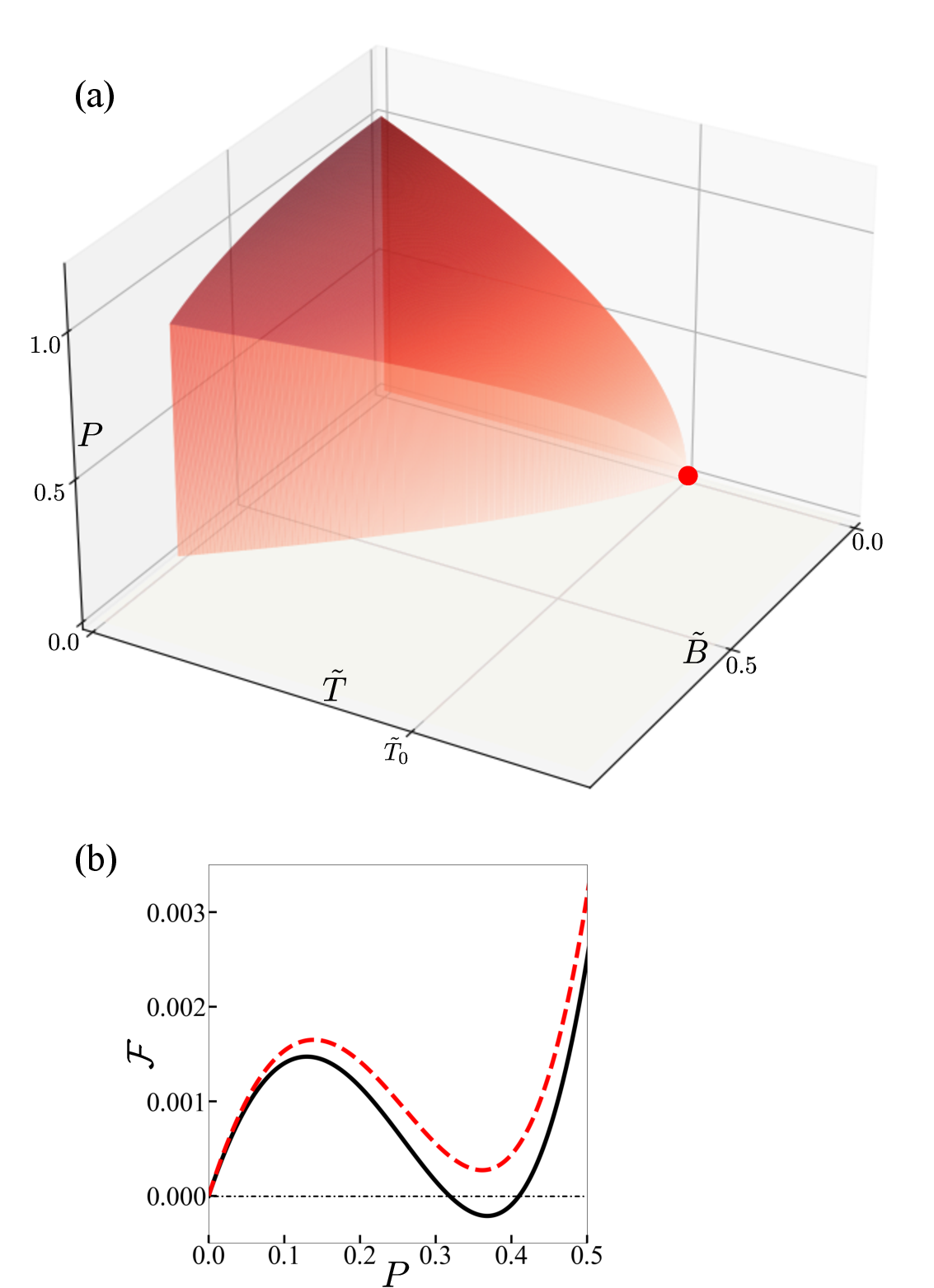}
\caption{
(a) The 3D
plot of polarization $P$ as a function of $\tilde{T}= \alpha T/\beta$
and $\tilde{B}$. We set $\tilde{T}_0= \alpha T_0 /\beta = 1$.
(b) The free energy $\mathcal{F}$ as
a function of $P$. We choose $\tau=0.2$.
The black solid represents $\tilde{B}=0.218$, while
the red dashed represents $\tilde{B}= 0.224$.
The true polarization for a given set of parameters is determined
by the global minimum of $\mathcal{F}$. }\label{fig1}
\end{figure}

Usually,
it is not easy to minimize the free energy because that depends on $\Psi(\textbf{r})$
which can be arbitrary function in real space.
Let us first try under a homogeneous assumption, $\nabla_i \Psi=0$
(i.e.,$\Psi$ being a constant).
In this case, the magnetoelectric coupling
between the magnetic field and the polarization becomes
$\left| \mathcal{A}_i \Psi\right|^2=B^2 P$,
which is independent of
the angle between the polarization ${\bf{P}}$ and magnetic field ${\bf{B}}$.
Such an independence is unnatural,
which will be removed after the spatial fluctuation of $\Psi(\textbf{r})$ is taken into account.
The free energy density (\ref{eq:freeF}) now becomes
\begin{equation}
\begin{split}\label{eq:unifreeenexp}
\mathcal{F} \approx & \frac{1}{2} \kappa^2 B^2 P + \left(T-T_0\right) \frac{\alpha}{2}
P^2 + \frac{\beta}{4} P^4,
\end{split}
\end{equation}
whose magnitude is of position-independent.
It is obvious that the free energy density~\eqref{eq:unifreeenexp}
is invariant under time reversal $\textbf{M}\to - \textbf{M}$
or space inversion $\textbf{P}\to - \textbf{P}$, as it should be for
a phenomenological theory of multiferroic materials.
Note that $P$ refers the module of the polarization whose value ranges from $0$ to $\infty$.

We find $\mathrm{d}\mathcal{F}/\mathrm{d}P > 0$ for arbitrary $P$
above the ferroelectric transition temperature ($T\geqslant T_0$), which means $\mathcal{F}$ increases monotonically
with $P$.
The minimization of $\mathcal{F}$ indicates $P\equiv 0$.
In other words, the net polarization is zero for whatever
magnetic field, a trivial result.
It becomes interesting if the temperature is below $T_0$.
For $T<T_0$, $\mathrm{d}\mathcal{F}/\mathrm{d}P$ is a convex function.
Next, we will
analyze the properties of $\mathcal{F}(P)$ and $\mathrm{d}\mathcal{F}/\mathrm{d}P$ to
obtain the minimum point of free energy. For convenience,
we introduce the rescaled magnetic field $\tilde{B}= \kappa B/\sqrt{\beta}$
and the rescaled temperature $\tilde{T} = \alpha T/\beta$.
And for $T<T_0$, the temperature difference from
critical temperature is denoted by $\tau =\tilde{T}_0-\tilde{T}\geqslant 0$.
Note that we only need to consider $B\geqslant 0$, since $\mathcal{F}$
depends on $B^2$ but is independent of the sign of $B$.

In the case of $\tilde{B} > \sqrt{\frac{4}{3\sqrt{3}}} \tau^{3/4}$,
we always have $\mathrm{d}\mathcal{F}/\mathrm{d}P > 0$ so that $\mathcal{F}$ increases monotonically,
the minimization of free energy leads to $P\equiv 0$.
For a smaller $\tilde{B}$, $\mathrm{d}\mathcal{F}/\mathrm{d}P$ intersects with the
positive $P$-axis twice.
As $P$ increases, the density of free energy $\mathcal{F}$ first increases to a local maximum and then decreases to a local minimum (the location is denoted as $P=P_\mathrm{c}$),
thereafter, it increases again till the infinity.
In Fig.~\ref{fig1}(b), we plot the dependence of $\mathcal{F}$ on $P$, respectively, for two
different values of magnetic fields $\tilde{B} = 0.218$ and $0.224$
where $\tau=0.2$ is fixed.
Note that these two values of $\tilde{B}$ are close to each other,
but the corresponding minimum points  differs dramatically.

To find the global minimum point,
we need to compare $\mathcal{F}(P_\mathrm{c})$ with $\mathcal{F}(P=0)=0$.
The defining equation of $P_\mathrm{c}$, i.e., $\frac{1}{\beta}\frac{d\mathcal{F}}{dP}|_{P=P_\mathrm{c}}=
P^3_\mathrm{c} - \tau P_\mathrm{c} + \tilde{B}^2/2 = 0$,
is a depressed cubic. Its root $P_\mathrm{c}$ can be obtained
by using the cubic formula.
Note that $P_\mathrm{c} $ is the largest root of the cubic equation.
After some straightforward calculations, we find
$4\mathcal{F}\left(P_\mathrm{c} \right)/\left(P_\mathrm{c} \beta\right) = 3\tilde{B}^2/2-\tau P_\mathrm{c} $.
Therefore, if $P_\mathrm{c} <3\tilde{B}^2/\left(2\tau\right)$, then $P=0$ is the global minimum
(see the red dashed in Fig.~\ref{fig1}(b)),
which means the polarization is zero. But if $P_\mathrm{c} >3\tilde{B}^2/\left(2\tau\right)$, then $P_\mathrm{c} $ is the
global minimum (see the black solid in Fig.~\ref{fig1}(b)),
which means the polarization is nonzero. It is easy to see that
the ferroelectric transition (between $P=0$ and $P=P_\mathrm{c} $)
must be first-order for $\tau >0$. For example, in Fig.~\ref{fig1}(b), as $B$ increases
a little bit from $0.218$ to $0.224$, the global minimum jumps from $P=0$
to $P_\mathrm{c}  \approx 0.35$.

We substitute $P=3\tilde{B}^2/\left(2\tau\right)$ into
the function $y= \frac{1}{\beta}\frac{\mathrm{d}\mathcal{F}}{\mathrm{d}P} = P^3 - \tau P + \tilde{B}^2/2$,
and then check whether $y(P=3\tilde{B}^2/\left(2\tau\right))$ is positive or negative.
Notice the fact that the function $y(P)$ is convex,
$P_\mathrm{c} $ is its largest zero point, and
the minimum of $y$ is located at $P=\sqrt{\tau/3}$.
After some calculations,
we find $P_\mathrm{c} <3\tilde{B}^2/\left(2\tau\right)$ when $\tilde{B}^4>
\left(2\tau/3\right)^3$, but
$P_\mathrm{c} >3\tilde{B}^2/\left(2\tau\right)$ when
$\tilde{B}^4 < \left(2\tau/3\right)^3$.
Therefore, the ferroelectric phase transition occurs at
\begin{equation}\label{eq:BTrelation}
\tilde{B} = \tilde{B}_\mathrm{c}\equiv \left( 2\tau /3\right)^{3/4}.
\end{equation}
For $\left| {\bf B}\right| <B_\mathrm{c} = \left( 2\tau /3\right)^{3/4} \sqrt{\beta}/\kappa$,
the polarization is nonzero, but
it becomes zero for $\left| {\bf B}\right| >B_\mathrm{c}$.
And such a transition is discontinuous except for $T=T_0$ at which $B_\mathrm{c}=0$.
As $T=T_0$, the transition is continuous.

In Fig.~\ref{fig1}(a), we plot the polarization as a function of temperature
and magnetic field. As $B=0$, the polarization changes continuously from
a finite value at $T<T_0$ to zero at $T\geqslant T_0$, which signifies
a continuous ferroelectric phase transition. But for $B>0$, the transition
occurs at $T_\mathrm{c} <T_0$, and it becomes discontinuous.
According to Eq.~\eqref{eq:BTrelation}, the relation
between the transition field strength and the transition temperature is given by
\begin{equation}
B_\mathrm{c}\propto \left(T_0 - T_\mathrm{c}\right)^{3/4}.
\end{equation}
This is a characteristic
feature of our theory. It remains valid even after the spatial fluctuation of polarization is taken into account.

\section{Spiral solution}
\label{sec:spiral}

Next we consider inhomogeneous case
where the polarization varies spatially,
i.e., generally proposing $\nabla_i \Psi \neq 0$.
We will look for a possible function $\Psi(\textbf{r})$ that minimizes
the free energy, $\int d^3\textbf{r} \mathcal{F}$.
Both our analytical analysis and numerical results
suggest that such $\Psi(\textbf{r})$ should display a spiral order.
Without loss of generality, we assume that the plane spanned
by the vector $\textbf{B}$ (in the $y$-direction) and the propagator vector of spiral order
is the $x$-$y$ plane.
Equivalently speaking, the propagator vector lies in the $x$-$y$ plane.
This means $\partial \Psi/\partial z = 0$,
i.e., the polarization keeps uniform in the $z$ direction.

Substituting $B_y = B$ and $B_x=B_z=0$ in (\ref{eq:freeF}),
we reexpress the free energy density as
\begin{equation}
\begin{split}\label{eq:inplaneF}
\mathcal{F} = &
\frac{1}{2} \kappa^2 B^2 \left|\Psi\right|^2 + \left(T-T_0\right) \frac{\alpha}{2}
\left|\Psi\right|^4 + \frac{\beta}{4} \left|\Psi\right|^8
 + \frac{1}{2}  \left|  \partial_x \Psi\right|^2 \\ &
+ \frac{1}{2} \left|  \partial_y \Psi\right|^2  +
\frac{\kappa B}{\sqrt{2}} \text{Im} \left[\psi_2 \partial_x \psi^*_2-
\psi_1 \partial_x \psi^*_1 \right].
\end{split}
\end{equation}
As in the absence of magnetic field $B=0$, the last term in above equation vanishes.
Thus a nonuniform $\Psi(\textbf{r})$ always causes the free energy to increase, because
$\left|\partial_x \Psi\right|^2$ and $\left|  \partial_y \Psi\right|^2$ are positively
definite.
As we have know in previous section, Sec.~\ref{sec:int},
the polarization favors a uniform solution for $B=0$.

\begin{figure}[tbp]
\includegraphics[width=0.8\linewidth]{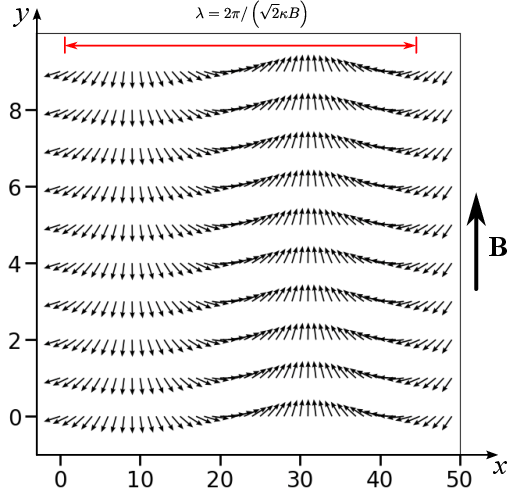}
\caption{Numerical solution of the polarization
distribution on a square lattice (lying in the $x$-$y$ plane) with the lattice
constant set to unit ($a=1$). The short black arrows represent
the polarizations at each lattice site. The magnetic
field $\textbf{B}$ is along the $y$-direction. We choose $\tilde{B}=0.1$,
$\tau = 5.0 $ and $\beta =1.0$. The red interval marker
has a length of $\lambda = 2\pi/\left( \sqrt{2}\kappa B \right)$, which
should be the spiral wave length according to our analytical analysis.
}\label{fig2}
\end{figure}

On the other hand, for novanishing $B$, a nonuniform solution is preferred
once if $\kappa B \text{Im} \left[\psi_2 \partial_x \psi^*_2-
\psi_1 \partial_x \psi^*_1 \right]< - \left( \left|  \partial_x \Psi\right|^2
+\left|  \partial_y \Psi\right|^2\right)/\sqrt{2}$.
This is possible if we assume,
for example, $\psi_2 \propto e^{i\kappa Bx/\sqrt{2}}$
and $\psi_1 \propto e^{-i\kappa Bx/\sqrt{2}}$.
Note that such $\psi_1$ and $\psi_2$ indicate
$P_x \propto \cos(\sqrt{2}\kappa B x)$ and $P_y \propto \sin(\sqrt{2}\kappa B x)$,
which describes a spiral-like electric polarization along the $x$-direction.
The spiral propagation direction is perpendicular to the direction of $\textbf{B}$.

Furthermore, we check whether a spiral solution has the lowest free energy.
Let us express the complex order parameters as
$\psi_1 = p \cos(\theta) e^{i\phi /2}$
and $\psi_2 = p \sin(\theta) e^{-i\phi /2}$.
Here we neglect the global phase as it can be simultaneously added to $\psi_1$ and $\psi_2$.
As global phase of $\Psi$ does not affect $\textbf{P}$ which has no explicit physical meaning,
one can neglect the global phase without loss of generality.
Here, $p^2 = \left| \psi_1\right|^2+
\left| \psi_2\right|^2 =P $ is the module of the polarization,
$\tan\theta$ is the ratio of $\left| \psi_2\right|$ to $\left| \psi_1\right|$,
and $\phi$ is the relative phase difference between $\psi_1$ and $\psi_2$.
With these notations, the last term of Eq.~\eqref{eq:inplaneF} becomes
$\text{Im} \left[\psi_2 \partial_x \psi^*_2-
\psi_1 \partial_x \psi^*_1 \right] = \frac{1}{2} P \partial_x \phi$, which depends
on the derivative of phase difference.
The presence of $\partial_x\phi$ is the key
reason that a non-uniform solution has lower free energy.
And $\partial_x\phi \neq 0$ will lead to spatial fluctuations
of $P_x$ and $P_y$ that take the form of trigonometric functions,
indicating a spiral order of polarization.

The total free energy density now reads
\begin{equation}
\begin{split}\label{eq:inFpsp}
\mathcal{F} =  &
\frac{1}{2} \kappa^2 B^2 P + \left(T-T_0\right) \frac{\alpha}{2}
P^2 + \frac{\beta}{4} P^4
\\ & + \frac{1}{2}  \left\{ \left(\partial_x p\right)^2 + p^2 \left(\partial_x \theta\right)^2
+ \left(\partial_y p\right)^2 + p^2 \left(\partial_y \theta\right)^2 \right. \\ & \left.
+ \frac{1}{4}p^2 \left(\partial_y \phi\right)^2 \right\}  +
\frac{p^2}{8} \left[ \left(\partial_x \phi\right)^2 +
\frac{4\kappa B}{\sqrt{2}} \partial_x \phi \right].
\end{split}
\end{equation}
To minimize $\mathcal{F}$, we demand that all the positive definite
terms (the terms in brace) be zero, that is $\partial_x p=
\partial_y p = \partial_x \theta = \partial_y\theta = \partial_y \phi =0$.
On the other hand, a nonzero $\partial_x \phi$ reduces $\mathcal{F}$.
According to the last term in Eq.~\eqref{eq:inFpsp}
(the terms in square bracket), the minimization requires
$\partial_x \phi=-\sqrt{2}\kappa B$, or
$\phi=-\sqrt{2} \kappa B x$.
Consequently, the free energy density reads
\begin{equation}\label{eq:nonunifree}
\mathcal{F} =
\frac{1}{4} \kappa^2 B^2 P + \left(T-T_0\right) \frac{\alpha}{2}
P^2 + \frac{\beta}{4} P^4.
\end{equation}
Equation~\eqref{eq:nonunifree} is similar to Eq.~\eqref{eq:unifreeenexp},
but with the magnetic field rescaled as $B\to B/\sqrt{2}$.
Obviously, $\mathcal{F}$ in Eq.~\eqref{eq:nonunifree}
is smaller than $\mathcal{F}$ in Eq.~\eqref{eq:unifreeenexp}
for arbitrary $B>0$. Therefore, a spiral solution
is always favored in the presence of magnetic field.

Since Eq.~\eqref{eq:nonunifree} can be regarded as Eq.~\eqref{eq:unifreeenexp}
with $B$ rescaled, the minimization of Eq.~\eqref{eq:nonunifree}
follows the same process as the minimization of Eq.~\eqref{eq:unifreeenexp},
while the latter has been discussed in Sec.~\ref{sec:int}.
As a result, there exists a transition field strength $\tilde{B}_\mathrm{c}$,
so that $\left| \textbf{P}\right|$ is finite for $\tilde{B}< \tilde{B}_\mathrm{c}$
but becomes zero for $\tilde{B}\geq \tilde{B}_\mathrm{c}$.
And the transition field strength must be
\begin{equation}
\tilde{B}_\mathrm{c} ={{\sqrt{2}}} \left(2\tau/3 \right)^{3/4}.
\end{equation}
In comparison with Eq.~\eqref{eq:BTrelation},
the ferroelectricity survives up to a larger magnetic field due to
the reduction of free energy caused by a spiral order.

To verify the existence of spiral order,
we carry out a pure numerical
algorithm to find the global minimum
of Eq.~\eqref{eq:inplaneF} on a square lattice. We start from
a randomly chosen initial configuration of $\Psi(\textbf{r})$,
and use the Powell method to find the global minimum of cost
function $\mathcal{F}$. The result is shown in Fig.~\ref{fig2}.
As can be seen, the polarization $\textbf{P}(x,y)$
does display a spiral pattern, confirming that the analytical analysis
is correct. 
The spiral wave propagates in the $x$-direction,
being perpendicular to the magnetic field. 
The wavelength of the spiral order is equal to
\begin{equation}
\lambda = 2\pi/\left(\sqrt{2}\kappa B\right),
\end{equation}
which coincides with $\phi=-\sqrt{2} \kappa B x$.

The polarization distribution in a magnetic field is now clear. For $B>B_\mathrm{c}$, there is
no electric polarization.
The destruction of polarization is caused by strong magnetic field.
For $B<B_\mathrm{c}$, the polarization always displays
a spiral order whose wavelength is inversely proportional to $B$.
In the limit $B\to 0$, the spiral wavelength becomes infinite,
and then we have a uniform polarization throughout the space.

\begin{figure}[tbp]
\includegraphics[width=0.8\linewidth]{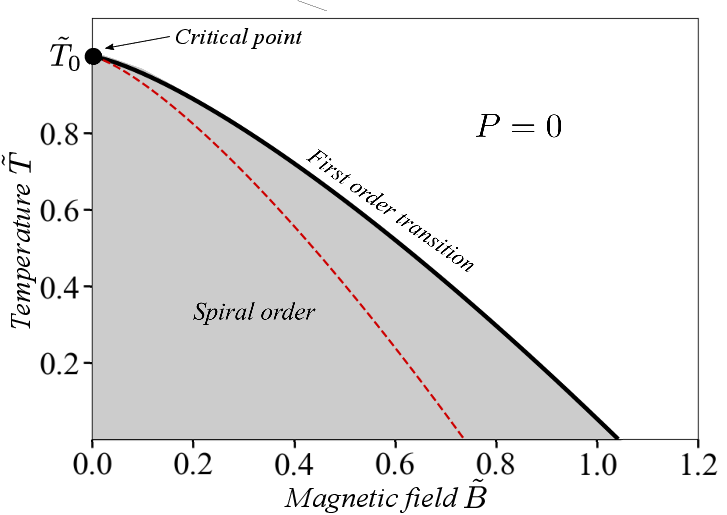}
\caption{The phase diagram in the $T$-$B$ space. We choose $\tilde{T}_0 =
\alpha T_0/\beta =  1$.
The grey area represents the spiral-order phase with finite polarization, while
the white area is the normal phase with $P=0$. These two phases
are separated by the solid line, at which a first-order transition happens
($P$ is discontinuous). As $\tilde{B}\to 0$, the transition line ends
at a critical point (black spot), at which the ferroelectric transition is continuous.
The red dashed represents the phase boundary obtained under the
homogeneous approximation, which is incorrect for neglecting
the spiral order.}\label{fig3}
\end{figure}

The overall phase diagram in the $T$-$B$ space is exhibited in Fig.~\ref{fig3},
which contains two different phases,
i.e. the spiral phase with finite polarizations and the
normal phase with $P=0$.
These two phases are separated by a first-order phase transition line, at which
the module of polarization changes discontinuously.
As the magnetic field goes to zero, the first-order transition line
approaches at a critical point where the ferroelectric transition becomes
continuous.

\section{Summary}
\label{sec:summ}

In this paper, we proposed a phenomenological model of magnetoelectricity in the multiferroic materials.
The characteristic feature of our model is a two-component complex order parameter that represents electric polarization,
and implies a direct coupling between polarization and magnetic field.
With increasing magnetic field strength,
we found that the absolute value of electric polarization continuously decreases until a transition point,
beyond which polarization abruptly drops to zero, signifying a first-order phase transition.
Our model successfully reproduces that a strong enough magnetic field will destroy the electric polarization.
And the transition field strength decreases with increasing
temperature, following a power law,
$B_\mathrm{c}\propto \left(T_0-T_\mathrm{c}\right)^{3/4}$
where $T_0$ is the ferroelectric transition temperature without a magnetic field.
Notably, the transition field strength $B_c$ is temperature-dependent.
At a weak magnetic field, the electric polarization exhibits a spiral order,
of which the propagation vector is perpendicular to $\textbf{B}$
and the wavelength is inversely proportional to $\left| \textbf{B}\right|$.
These predictions is expected to be  tested in future experiments on multiferroic materials.

\end{document}